\documentclass[12pt]{article}
\usepackage{epsfig}

\newcommand{\be}{\begin{equation}}
\newcommand{\dd}{\displaystyle}
\newcommand{\ee}{\end{equation}}
\newcommand{\bea}{\begin{eqnarray}}
\newcommand{\eea}{\end{eqnarray}}

\newcommand{\de}{\partial}
\begin{document}
\hfill$\vcenter{
 \hbox{\bf BARI-TH 426/01}
\hbox{\bf FIRENZE-DDF-382/01/02}
 \hbox{\bf UGVA-DPT-2001 11/1097} }$
\begin{center}
{\Large\bf\boldmath {Anisotropy Parameters for the Effective
Description of Crystalline Color Superconductors}}
\\ \rm \vskip1pc {\large
R. Casalbuoni$^{a,b}$,  R. Gatto$^c$, M. Mannarelli$^{d,e}$ and\\
G. Nardulli$^{d,e}$}\\ \vspace{5mm} {\it{$^a$Dipartimento di
Fisica, Universit\`a di Firenze, I-50125 Firenze, Italia
\\
$^b$I.N.F.N., Sezione di Firenze, I-50125 Firenze, Italia\\
$^c$D\'epart. de Physique Th\'eorique, Universit\'e de Gen\`eve,
CH-1211 Gen\`eve 4, Suisse\\ $^d$Dipartimento di Fisica,
Universit\`a di Bari, I-70124 Bari, Italia  \\$^e$I.N.F.N.,
Sezione di Bari, I-70124 Bari, Italia }}
\end{center}
%%% ----------------------------------------------------------------------

\begin{abstract}
In the high density low temperature limit, Quantum Chromodynamics
(QCD) exhibits a transition to a phase characterized by color
superconductivity and by energy gaps in the fermion spectra. Under
specific circumstances the gap parameter  has a crystalline
pattern, breaking translational and rotational invariance. The
corresponding phase is the the crystalline color superconductive
phase (or {\it LOFF} phase). In the effective theory the fermions
couple to the phonon arising from the breaking of rotation and
translation invariance. We compute the parameters of the low
energy effective lagrangian describing the motion of the free
phonon in the high density medium and derive the phonon dispersion
law.
\end{abstract}

\vskip1.cm
\section{Introduction}
QCD at large densities is expected to exhibit color
superconductivity  \cite{review} from condensation of quark
bilinears. It has recently been pointed out
\cite{LOFF,LOFF2001,LOFF5,LOFFbis,LOFF7,LOFF8,LOFF_rob,LOFF9,LOFF10}
%\cite{LOFF},
%\cite{LOFF2001}, \cite{LOFF5}, \cite{LOFFbis}, \cite{LOFF7},
%\cite{LOFF8}, \cite{LOFF_rob}, \cite{LOFF9}, \cite{LOFF10}
that a distinctive phase, called crystalline color superconducting
phase, may appear, in situations where different quark flavors
have different chemical potentials, with potential differences
lying inside a certain window.  Crystalline color
superconductivity is also expected to occur simply because of the
quark mass differences \cite{massdiff}. However in the following
we shall limit the discussion to the case of two massless quarks
with different chemical potentials.

Besides its theoretical interest for the study of the phase
structure of QCD theory, such a crystalline phase may result
relevant  for astrophysical dense systems \cite{LOFF},
\cite{LOFF8}, \cite{LOFF3}. A pairing ansatz similar to that of
the QCD crystalline phase was already encountered in condensed
matter (and referred to as LOFF phase from the name of the authors
who studied it in superconductivity \cite{LOFF2}). Other recent
applications of the LOFF phase in organic superconductors, nuclear
matter and cold trapped gases are respectively in Refs.
\cite{condmat}, \cite{nuclearmatter} and \cite{coldtrapped}.

In the crystalline phase, rotation and translation invariance are
both spontaneously broken. The ensuing goldstone structure and the
effective lagrangian were studied in \cite{LOFF5}. For plane-wave
behavior of the crystalline condensate the conclusion was that a
single goldstone appears, the phonon. The appearance and
description of a single goldstone in a situation where both
translations and rotations are spontaneously broken was found in
\cite{LOFF5} to originate from a functional relation to be
satisfied by fictitious fields associated to broken rotations such
as to assure small long wavelength fluctuations of the order
parameters.

The main purpose of this paper is to calculate at first order in
the derivative expansion the parameters of the effective
lagrangian for the crystalline phase and derive the phonon
dispersion law. Due to the fact that quark condensation singles
out one direction in space, the phonon dispersion law is
anisotropic.

The paper is organized as follows. In Section \ref{sec2} we review
the velocity-dependent effective lagrangian approach we shall use
in this paper. In Section \ref{sec3}  we review the basic facts
concerning the crystalline super-conducting phase and the
modifications that are needed in the effective lagrangian to deal
with the anisotropy induced
 by the LOFF condensates. In Section \ref{2.7} we derive the gap equations
for the BCS and the LOFF condensates by
 a Schwinger Dyson equation derived in the framework
 of the effective theory. In Section \ref{sec5} we introduce the phonon
field associated
 with the breaking of rotational and translational invariance and we write
down its coupling to the fermion quasiparticles.
In Section \ref{sec6} we derive the parameters of the effective
lagrangian for the phonon. Finally in Section \ref{secconc} we
draw our conclusions.
\section{Review of the velocity-dependent effective lagrangian\label{sec2}}
We shall employ in the sequel the effective lagrangian approach
based on velocity-dependent fields. It has been developed in
\cite{hong,beane,gatto,mannarelli}
 for the 2SC and CFL models and in \cite{LOFF5} for the LOFF
 phase. In this Section we give a brief review of it.
Let us consider the fermion field
 \be \psi(x)=\int\frac{d^4 p}{(2\pi)^4}e^{-ip\cdot x}\psi(p)\ee
 and  decompose the fermion momentum as
 \be p^\mu=\mu v^\mu+\ell^\mu\label{eq:2}\ee with
$v^\mu=(0,\vec v)$ and $\vec v$ the Fermi velocity (for massless
fermions $|\vec v|=1$); finally $\ell^\mu$ is a residual momentum.

By the decomposition (\ref{eq:2}) only the positive energy
component $\psi_+$ of the fermion field survives in the lagrangian
in the $\mu\to\infty$ limit  while the the negative energy
component $\psi_-$ can be integrated out. These effective fields
are velocity dependent and are related to the original fields by
\be\psi(x)=\sum_{\vec v}e^{-i\mu v\cdot x
}\left[\psi_+(x)+\psi_-(x)\right]\label{decomposition}
 ,\ee where
\be\psi_\pm(x)=\frac{1\pm \vec \alpha\cdot \vec v }2 \int
\frac{d\vec\ell}{(2\pi)^3} \int_{-\infty}^{+\infty}
\frac{d\ell_0}{2\pi} \, e^{-i\ell\cdot x} \psi_{\vec v}(\ell)\
.\ee Here $\displaystyle\sum_{\vec v} $ means an average over the
Fermi velocities  and \be\psi_\pm(x)\equiv\psi_{\pm,\vec v}(x) \ee
are velocity-dependent fields. The positive energy effective field
carries color and flavor indices: $\psi_+\equiv \psi_{+ ,i\,
\alpha}$ where $\alpha=1,2,3$ is a color index and $i$ is a flavor
index . In this paper $i=1,2$ (two flavors).
 The integration measure in the isotropic (CFL or two-flavor
2SC models) is as follows \be \int \frac{d\vec\ell}{(2\pi)^3}=\int
\frac{d\vec\ell_\perp}{(2\pi)^3}\int_{-\delta}^{+\delta}d\ell_\parallel=
frac{\mu^2}{(2\pi)^3} \int_0^{2\pi}d\phi\int_{-1}^{+1}d\cos\theta
\int_{-\delta}^{+\delta}d\ell_\parallel \ee where
$\ell_\parallel$ is along $\vec v$ and $d\vec\ell_\perp$ is a
surface element, which on a sphere of radius $\mu$ is equal to
$\mu^2\,d\Omega\,=\,\mu^2\, d\phi\, d\cos\theta$;
$\ell_\parallel$ is limited by $|\ell_\parallel |\le\delta$,
where $\delta$ is such that ($\Delta$ the BCS gap) \be
\Delta\ll\delta\le\mu \ .\label{delta}\ee The condition
$|\ell_\parallel |\le\delta$ is equivalent to the requirement \be
||\vec p|-\mu|\le\delta\ee once one recognizes that, due to the
spherical symmetry, one can always choose $\vec\ell$ in the
direction of $\vec v$. On the other hand the condition
(\ref{delta}) has a twofold implication: $\delta\le\mu$ means that
one is considering only degrees of freedom near the Fermi surface,
while, by the choice $\Delta\ll\delta$, the physical quantities
become independent of the cut-off procedure.

In the case of anisotropic superconductivity, such as the LOFF
phase, the situation is different and will be discussed in the
next section.

 It is useful to use a different basis for the fermion fields
 by writing \bea \psi_{+,i\alpha }&=&
\sum_{A=0}^3\frac{(\sigma_A)_{i\alpha}}{\sqrt 2}\varphi_{+}^A
~~~~~~~~(i,\,\alpha=1,\,2)~\cr \psi_{+,13}&=&\varphi_{+}^4\cr
\psi_{+,23}&=&\varphi_{+}^5\ ,\eea where $\sigma_A$ are the Pauli
matrices for $A=1,2,3$ and $\sigma_0=1$. $\varphi_{+}^A$ are
positive energy, velocity dependent fields: \be
\varphi_{+}^A\equiv\varphi_{+,\vec v}^A\ , \ee and represent
left-handed Weyl fermions. In view of the introduction of a gap
term (see next section) we introduce also the fields  \be
\varphi_{-}^A\equiv\varphi_{+,-\vec v}^A\ .\ee $\varphi_{\pm }^A$
should not be confused with the positive and negative energy
fields; they are both positive energy fields, but are relative to
opposite velocities.

The effective lagrangian $ {\mathcal L}_0$ for gapless fermion
fields at high density can therefore be written as follows
\footnote{Here $V^\mu=(1,\,\vec v)$ while $\tilde V^\mu=(1,\,-\vec
v)$. The derivation of (\ref{l0}) is in \cite{gatto}. It can be
obtained noting that, in the $\mu\to\infty$ limit, terms bilinear
in the fermion fields must have the same velocity by the
Riemann-Lebesgue lemma, while the spin matrices are substituted by
the $V^\mu$ vector by making use of  simple algebraic identities
\cite{hong}.}:
 \be {\mathcal L}_0\ =\ \sum_{\vec v}\sum_{A=0}^5
 \varphi_{+}^{A\dag}( i V\cdot\partial)\,\varphi_{+}^A  \ +\ (L\to
R)\ .\label{l0} \ee Using the fact that the average over
velocities is symmetric we write:
 \be{\mathcal L}_0 =
 \frac 1 2
  \sum_{\vec v}\sum_{A=0}^5\Big
 (\varphi_{+}^{A\,\dag} (i
V\cdot\partial)\, \varphi_{+}^A+ \varphi_{-}^{A\,\dag}( i \tilde
V\cdot\partial)\,\varphi_{-}^A\Big)\ .\label{eq:10}\ee
 Introducing now
\begin{equation}
\chi^A\ =\frac 1{\sqrt{2}}\left(\begin{array}{c}
  \varphi_{+}^A  \\ {}\\
  C\varphi_{-}^{A\,*}\ ,
\end{array}\right)
\end{equation}
the lagrangian can be written as follows: \be {\mathcal L}_0\ =\
\sum_{\vec v}\sum_{A=0}^5 \chi^{A\,\dag}\left(
\begin{array}{cc}
 i V\cdot\partial\ & 0
\\
0 & i \tilde V\cdot\partial
\end{array}
\right)\chi^A\ .\ee
 To this lagrangian involving only the left
  handed fields, one should add a similar one containing the right
  handed fermionic fields. Also notice that we have
  embodied the factor
$1/2$ appearing in Eq. (\ref{eq:10}) in the definition of the sum
over the Fermi velocities. This extra factor cannot be eliminated
by a redefinition of the fields $\psi_\pm$ since it corresponds to
a genuine doubling of the degrees of freedom; its appearance is
needed because, by introducing the fields with opposite $\vec v$,
$\psi_-=\psi_{+,-\vec v}$ we must integrate only over half solid
angle\footnote{In \cite{gatto} another factor $1/2$ appears in the
lagrangian ${\cal L}_D$; here we simplify the notation and get rid
of it by a redefinition of the fermion fields.}.

\section{Crystalline colour superconductive phase\label{sec3}} As shown in
\cite{LOFF}  when the two fermions have different chemical
potentials $\mu_1\neq \mu_2$, for $\delta\mu$ of the order of the
gap\footnote{The actual value of the range compatible with the
presence of the LOFF state depends on the calculation by which the
crystalline colour state is computed. Assuming a local interaction
as in \cite{LOFF} produces a rather small interval, which is
enlarged by  assuming gluon exchange, as in \cite{LOFFbis}.} the
vacuum state is characterized by a non vanishing expectation value
of a quark bilinear which  breaks translational and rotational
invariance. The appearance of this condensate is a consequence of
the fact that for $\mu_1\neq\mu_2$, and in a given range of
$\delta\mu=|\mu_1-\mu_2|$, the formation of a Cooper pair with a
total momentum\be \vec p_1\,+\,\vec p_2\,=\,2\vec q\neq \vec 0\ee
 is energetically favored in comparison with the normal
BCS state. The analysis of \cite{LOFF} shows that two different,
but related condensates are possible, one with the two quarks in a
spin zero state (scalar condensate) and the other one
characterized by total spin 1 (vector condensate). In the BCS
state the quarks forming the Cooper pair have necessarily $S=0$;
as a matter of fact, since the quarks have opposite momenta and
equal helicities, they must be in an antisymmetric spin state.
This is not true if the total momentum is not zero and the two
quarks can have both $S=0$ and $S=1$.

The possible form of these condensates is discussed in
\cite{LOFF}; these authors assume only two flavors and make the
ansatz of a plane wave behavior for both condensates: $\dd\propto
e^{2i\vec q\cdot\vec x}$. Though more complicated structures are
possible we will make the same  hypotheses. The results of
\cite{LOFF} are as follows: \be
-<0|\epsilon_{ij}\epsilon_{\alpha\beta 3 } \psi^{i\alpha}( \vec
x)C\psi^{j\beta}(\vec x)|0>= 2\Gamma_A^L e^{2i\vec q\cdot\vec
x}\label{scalar}\ ;\ee besides the scalar condensate
(\ref{scalar}) one may have a spin 1 vector condensate, symmetric
in flavor: \be i<0|\sigma^1_{ij}\epsilon_{\alpha\beta 3 }
\psi^{i\alpha}(\vec x)C\sigma^{03}\psi^{j\beta}(\vec
x)|0>=2\Gamma_B^L e^{2i\vec q\cdot\vec x}\ .\label{vector}\ee The
effect of the two non vanishing vacuum expectation values can be
taken into account by adding to the lagrangian the term: \bea
{\cal L}_\Delta&=&{\cal L}_\Delta^{(s)}+{\cal
L}_\Delta^{(v)}=\cr&&\cr &=&-\frac{ e^{2i\vec q\cdot\vec x}
}2\,\epsilon^{\alpha\beta 3}
\psi_{i\alpha}^T(x)C\left(\Delta^{(s)} +\vec\alpha\cdot\vec n
\Delta^{(v)} \sigma^1_{ij}\right) \psi_{i\beta}(x)\,+\,{\rm h.c.}\
, \label{ldeltaeff}\eea which includes both the scalar and the
vector condensate.

 Let us consider the lagrangian term relative to
the scalar condensate. We shall write it  as follows: \be {\cal L
}^{(s)}_\Delta=-\frac{\Delta^{(s)}}2\, e^{2i\vec q\cdot\vec x}
\epsilon^{\alpha\beta 3}\epsilon_{ij} \psi_{i\alpha}^T(x)C
\psi_{i\beta}(x)\ -(L\to R)+{\rm h.c.} \label{loff5}~.\ee Here
$\psi(x)$ are positive energy left-handed fermion fields discussed
in the previous section. We neglect the negative energy states,
consistently with the assumption discussed in the previous
section.

 In order to introduce velocity dependent positive energy
fields $\psi_{+,\,\vec v_i;\,i\alpha}$ with flavor $i$ we have to
decompose the  fermion momenta according to (\ref{eq:2}): \be \vec
p_j=\mu_j\vec v_j+{\vec \ell}_j \hskip1cm (j=1,2)\ .
\label{dec1}\ee Therefore we have : \bea {\cal L
}^{(s)}_\Delta&=&-\frac{\Delta^{(s)}} 2 \,\sum_{\vec v_1,\vec v_2}
\exp\{i\vec x\cdot\vec\alpha(\vec v_1,\,\vec v_2,\,\vec q
)\}\epsilon_{ij}\epsilon^{\alpha\beta 3}\psi_{+,\,-\,\vec
v_i;\,i\alpha}(x)C \psi_{+,\,-\,\vec v_j;\,j\beta}(x)\cr && -(L\to
R)+{\rm h.c.}\ ,\label{loff6}\eea where \be\vec\alpha(\vec
v_1,\,\vec v_2,\,\vec q)=2\vec q-\mu_1\vec v_1-\mu_2\vec v_2\ .\ee
We choose
 the $z-$axis along $\vec q$ and the vectors $\vec v_1,\,
 \vec v_2$ in the $x-z$ plane; if $\alpha_1,\,\alpha_2$ are the angles
 formed by the vectors $\vec v_1,\,\vec v_2$ with the $z-$axis we have
\bea \alpha_x&=&\
-\mu_1\sin\alpha_1\cos\phi_1-\mu_2\sin\alpha_2\cos\phi_2\ ,\cr
\alpha_x&=&\
-\mu_1\sin\alpha_1\sin\phi_1-\mu_2\sin\alpha_2\sin\phi_2\ ,\cr
\alpha_z&=&\ 2\,q\, -\,\mu_1\cos\alpha_1\,-\,\mu_2\cos\alpha_2 \
.\label{alfaz}\eea Because of the already quoted Riemann-Lebesgue
lemma, in the $\mu_1,\,\mu_2\,\to\infty$ limit the only non
vanishing terms in the sum correspond to the condition \be
\alpha_x= \alpha_y=\ 0\ .\label{4bis}\ee Let us introduce
\bea\mu&=&\frac{\mu_1+\mu_2}{2}\cr\delta\mu&=&\,-\,\frac{\mu_1-\mu_2}{2}
.\label{dec2}\eea Eq. (\ref{4bis}) implies
\be\alpha_1=\alpha_2+\pi+{\cal O}\left(\frac 1 \mu\right)\ ,\ee
i.e. the two velocities are opposite in this limit.

The condition on $\alpha_z$ depends on the behavior of $q$ in the
limit $\mu\to\infty$. If we might take the limit $q\approx
\delta\mu\to\infty$ as well, we would also have $\alpha_z\approx
0$; we will make this approximation which is justified by previous
analysis \cite{LOFF}, which points to a value of $q\simeq 1.2
\delta\mu$ and $\delta\mu\sim 0.7\Delta_{BCS}\gg \Delta^{(s,v)}$.
These conditions have to be justified also in our approach and we
will do it in the next section. Notice that this approximation
considerably simplifies the loop calculation in presence of the
LOFF condensate, which is usually performed by defining a blocking
region, where the LOFF condensate is impossible, and an allowed,
pairing region, whose definition is mathematically involved. This
approximation may be therefore useful in complex calculations such
as the present one, even though it can produce numerical results
somehow different by the more precise calculations given in
\cite{LOFF}. We have:\bea
 \cos\alpha_1\,&=&\,-\,\frac{\delta\mu}{q}+\frac{q^2-\delta\mu^2}{q\mu}
 \cr
 \cos\alpha_2\,&=&\,+\,\frac{\delta\mu}{q}+\frac{q^2-\delta\mu^2}{q\mu} \
 ,\label{28}
 \eea which, together with (\ref{4bis}), has the solution
\bea \alpha_2&\ \equiv\ &\theta_q\ =\
\arccos\frac{\delta\mu}q\,-\,\frac\epsilon 2\ ,
\label{tetaq}\\
\alpha_1&\ =\ &\alpha_2+\pi-\epsilon\eea with
\be\epsilon\,=\,2\,\frac{\sqrt{q^2-\delta\mu^2}}\mu \ .\ee
Therefore, as anticipated, $\dd\epsilon={\cal O}\left(\frac 1
\mu\right)$ and in the limit $\mu\to\infty$ the two velocities are
almost
 antiparallel:\be\vec v_1\simeq -\vec v_2\ .\ee
Putting
 \be \psi_{+,\,\pm\vec
v_i;\,i\alpha}(x)~\equiv~\psi_{\pm\vec v_i;\,i\alpha}(x)~,
 \ee eq. (\ref{loff6}) becomes: \be{\cal L
}^{(s)}_\Delta=-\frac{\Delta^{(s)}}{2} \sum_{\vec
v}\epsilon_{ij}\epsilon^{\alpha\beta 3} \psi_{+\vec
v;\,i\alpha}(x)C \psi_{-\vec v;\,j\beta}(x)-(L\to R)+{\rm
h.c.}\label{eq:20}\ee In a similar way the term corresponding to
the vector condensate in the lagrangian
  can be written as follows: \be {\cal L
}^{(v)}_\Delta=\, - \frac{\Delta^{(v)}}2 \sum_{\vec
v}\sigma^1_{ij}\epsilon^{\alpha\beta 3} \psi_{+\vec
v;\,i\alpha}(x)C(\vec v\cdot\vec n ) \psi_{-\vec v;\,j\beta}(x)
-(L\to R)+{\rm h.c.} \label{eq:23} \ee where $\vec n=\vec q/|\vec
q|$  is the direction corresponding to the total momentum carried
by the Cooper pair and we have used $\psi_-^TC\vec\alpha\cdot \vec
n\psi_+=\vec v\cdot\vec n\psi_-^TC\psi_+$~.

The procedure of the limit $q\to\infty$ may produce a finite
renormalization of the gap parameters. To show this let us
consider in more detail how  the second of the two equations
(\ref{28}) can be obtained. In order to take the limit in the
lagrangian for $\mu\to\infty$ or $q\to\infty$ we perform a
smearing in the space time integrations; for example, considering
the $z$ axis we introduce   a smearing $|\Delta z|$ and we write:
\begin{eqnarray}
 e^{i2qhz}&\to& \frac{1}{|\Delta z|}
 \int_{z-|\Delta z|/2}^{z+|\Delta z|/2}e^{i2qhy} dy
 =e^{i2qhz}\,\frac{\sin(qh|\Delta z|)}{qh|\Delta z|}=
\cr &=&  \frac{\pi}{R}e^{i2qhz}\,\delta_R(h)\end{eqnarray}
  We have put \be
  R=q|\Delta z|\ee and introduced the "fat delta" $\delta_R(x)$  defined by
 \be
\delta_R(x)\equiv\frac{\sin(Rx)}{\pi x} \ ,\label{tre}\ee which,
for large $R$, gives \be\delta_R(x)\to \delta(x) \ .\ee Let us now
justify the approximation (\ref{28}). We have ($\theta_2=$
azimuthal angle of the velocity vector $\vec v_2$) \bea \int
d\cos\theta_2 e^{i2qh(\cos\theta_2)z}&\rightarrow&
\frac{2\pi}{2q|\Delta z|}\int d\cos\theta_2
e^{i2qh(\cos\theta_2)z}\delta_R\left(h(\cos\theta_2)\right)\cr
&\approx& \frac{\pi}{q|\Delta z||h^\prime(\cos\alpha_2)|}\int
d\cos\theta_2\delta(\cos\theta_2-\cos\alpha_2 )
 \ . \label{a2}\eea In the previous equations we
have defined\footnote{The reader should note that, by our
convention, $\delta\mu<0$.} \be
h(\cos\theta_2)=1+\frac{1}{2q}\left(-\mu_2\cos\theta_2+
\cos\theta_2\sqrt{\mu^2_2-\frac{4\mu\delta\mu}{\cos^2\theta_2}}
\right)\ .\ee In the sequel we will also use \be
h_\pm(\cos\theta_2)=1+\frac{1}{2q}\left(-\mu_2\cos\theta_2\pm\sqrt{\mu^2_2\cos^2
\theta_2-4\delta\mu\mu} \right)\ \ee where one has $\pm
1=+sign(\cos\theta_2)$.  Due to eq. (\ref{a2}),  the condensates
are modified as \be \Delta \to \frac \pi
R\delta_R(h_\pm(cos\theta))\, \Delta\ .\ee We handle the fat
delta according to the Fermi trick in the Golden Rule; in
expressions involving the gap parameters we make one substitution
\be \delta_R(h_\pm(cos\theta))\to\delta(h_\pm(cos\theta)) \ee in
the numerator;  correspondingly we will get a factor \be
k_R=\frac{\pi|\delta\mu|}{qR}\ee in the numerator (from the region
$cos\theta<0$); the other fat delta are computed as follows
 \be
\frac{\pi\delta_R(h_-(cos\theta))} R\to\frac{\pi\delta_R(0)}R\to
\,1 .\ee
 In
the basis introduced in the previous section, the effective
lagrangian is
 \bea
  {\mathcal L}&=& {\mathcal L}_0 \ +\ {\mathcal L}^{(s)}_\Delta \
  +\ {\mathcal L}^{(v)}_\Delta\ =\cr&&\cr
&=& \sum_{\vec v}\sum_{A=0}^5 \chi^{A\,\dag}\left(
\begin{array}{cc}
 i\, \delta_{AB}\ V\cdot\partial\ & \,\Delta_{AB}^\dag
\\
\,  \Delta_{AB} & i\,\delta_{AB}\ \tilde V\cdot\partial\
\end{array}\right)\chi^B\ .\eea
The matrix $\Delta_{AB}$ is as follows: \be
\Delta_{AB}=0\hskip1cm(A\,{\rm or} \,B=4\,{\rm or} \,5 )\ee while
for $A,B=0,...,3$ we have:\be \Delta_{AB}=\frac \pi
R\delta_R(h_\pm(cos\theta)\, \left(\begin{array}{cccc}
  \Delta_0 & 0 & 0 & -\Delta_1 \\
  0 & -\Delta_0 &- i\Delta_1 & 0 \\
  0 & +i \Delta_1&-\Delta_0& 0 \\
  \Delta_1 & 0 & 0 & -\Delta_0
\end{array}\right)\ ,\label{eq:38}
\ee with \bea \Delta_0\,&=&\, \Delta^{(s)}\cr \Delta_1\,&=\,&\vec
v\cdot\vec n\, \Delta^{(v)}\ .\eea

Let us now discuss the precise meaning of the average over
velocities. As discussed above we embodied a factor of 1/2
 in the
average over velocities; however in the LOFF case the sum over the
velocities is no longer symmetric and, whenever the integrand
contains the matrix $\Delta_{AB}$ now reads: \be \sum_{\vec
v}\equiv \ \sum_{\vec v}\frac \pi R\delta_R(h_\pm(cos\theta)\,\
\to \frac{k_R}{2}\int\frac{d\phi}{2\pi}\ , \label{sum3}\ee if
\be\vec v=\left(\sin\theta_q\cos\phi,\,
\sin\theta_q\sin\phi,\,\cos\theta_q\right)\ ,\ee
 and $\theta_q$ given in (\ref{tetaq}). As discussed above,
 this limit corresponds  to $R$ large.
  For future reference we observe that, in the same limit,
\be \sum_{\vec
v}\,v_iv_j\,=\,k_R\,\left(\frac{\sin^2\theta_q}{4}\,
\left(\delta_{i1}\delta_{j1}+\delta_{i2}\delta_{j2}\right)
\,+\,\frac{\cos^2\theta_q}{2}\delta_{i3}\delta_{j3} \right)\ .\ee

The effective action for the fermi fields in momentum space reads:
\be S=\sum_{\vec v}\sum_{A,B=0}^5 \int \frac{d^4\ell}{(2\pi)^4}
\frac{d^4\ell^\prime}{(2\pi)^4}
\chi^{A\dag}(\ell^\prime)D^{-1}_{AB}(\ell^\prime,\ell)\chi_B(\ell)~,
 \ee
 where $D^{-1}_{AB}(\ell^\prime,\ell)$ is the inverse propagator,
 given by:
\be D^{-1}_{AB}(\ell^\prime,\ell)\,=\,\left(
\begin{array}{cc}
V\cdot\ell  \delta_{AB}& \Delta_{AB}^\dag
\\
\Delta_{AB}
 &  \tilde V\cdot\ell
 \delta_{AB}
\end{array}\nonumber
\right)\,\delta^4(\ell^\prime-\ell)\ .\ee From these equations one
can derive the quark propagator, defined by\be \sum_B \int
\frac{d^4\ell^\prime}{(2\pi)^4} D^{-1}_{AB}(\ell,\ell^\prime)
D_{BC}(\ell^\prime,\ell^{\prime\prime})\,=\,\delta_{AC}
\delta^4(\ell\,-\, \ell^{\prime\prime}) \ .\ee It is given by \be
\displaystyle D_{AB}(\ell,\ell^{\prime\prime})=
(2\pi)^4\delta^4(\ell-\ell^{\prime\prime})\times\sum_C\left(
\begin{array}{cc} \displaystyle
 \frac{ \tilde V\cdot\ell\,\delta_{AC}}{\tilde D_{CB}(\ell)}\,
& \displaystyle -\frac{\Delta_{AC}^\dag}{D_{CB}(\ell)}
\\ \\
\displaystyle -\frac{\Delta_{AC}}{\tilde D_{CB}(\ell)}
 &    \displaystyle
 \frac{ V\cdot\ell\,\delta_{AC}}{D_{CB}(\ell)}\,
\end{array}
\right)\label{propagatore}\ee where \bea &D_{CB}(\ell)&=
\left(V\cdot \ell\,\tilde V\cdot
\ell\,-\,\Delta\Delta^\dag\right)_{CB}\cr
 &\tilde D_{CB}(\ell)&=  \left(V\cdot \ell\,\tilde
V\cdot \ell\,-\,\Delta^\dag\Delta\right)_{CB}\ . \eea

 The propagator for  the fields $\chi^{4,5}$
does not contain gap mass terms and is given by \be
D(\ell,\ell^{\prime})=(2\pi)^4\,\delta^4(\ell-\ell^{\prime})\,\left(
\begin{array}{cc}
 (V\cdot\ell)^{-1}&0\\
 0&  (\tilde V\cdot\ell)^{-1}
\end{array}
\right)~.\ee For the other  fields $\chi^A$, $A=0,\cdots,3$, it is
useful to go to a representation where $\Delta\Delta^\dag$ and
$\Delta^\dag\Delta$ are diagonal. It is accomplished by performing
a unitary transformation which transforms the basis $\chi^A$ into
the new basis $\tilde \chi^A$ defined by \be
\tilde\chi^{A}=R_{AB}\chi^{B}\ ,\label{newbasis1}\ee with \be
R_{AB}=\frac 1 {\sqrt 2 } \left(\begin{array}{cccc}
  1 & 0 & 0 & 1 \\
  0 & 1 &- \,i & 0 \\
  0 & +i &-\,1& 0 \\
 1 & 0 & 0 & -\,1
\end{array}\right)\ .\label{newbasis2}
\ee In the new basis we have
 \bea
\left(\Delta\Delta^\dag\right)_{AB}&=&\lambda_A\delta_{AB}\cr
\left(\Delta^\dag\Delta\right)_{AB}&=&\tilde\lambda_A\delta_{AB}\eea
where\bea \lambda_A&=&\left(\frac \pi
R\delta_R(h_\pm(cos\theta)\,\right)^2((\Delta_0+\Delta_1)^2,
\,(\Delta_0-\Delta_1)^2, \,(\Delta_0+\Delta_1)^2,
\,(\Delta_0-\Delta_1)^2 )\cr&&\cr \tilde\lambda_A&=&\left(\frac
\pi
R\delta_R(h_\pm(cos\theta)\,\right)^2\left((\Delta_0-\Delta_1)^2,
\,(\Delta_0-\Delta_1)^2, \,(\Delta_0+\Delta_1)^2,
\,(\Delta_0+\Delta_1)^2\right)\cr&&\ .
 \eea For further reference we also define
  \be \mu_C=(\Delta_0+\Delta_1, \,\Delta_1-\Delta_0,
\,\Delta_0+\Delta_1, \,\Delta_1-\Delta_0) \label{muc} \ .\ee

Let us finally discuss the integration limits in the residual
momentum $\vec \ell$ in the LOFF case.  We write:
\be\vec\ell\,=\,\vec\ell_\perp\,+\,\vec\ell_\parallel \ ,\ee where
$\vec\ell_\parallel$ is parallel to $\vec v_2\simeq\,-\,\vec v_1$
and   $\vec\ell_\perp$ is a 2-dimensional vector orthogonal to
$\vec v_2$. Clearly $V\cdot\ell=\ell_0-\ell_\parallel$ and $\tilde
V\cdot\ell=\ell_0+\ell_\parallel$, therefore the propagator only
depends on $\ell_0$ and $\ell_\parallel$; however, since the
integration over velocities is anisotropic, the theory,
differently from the CFL and 2SC cases, is not 2-dimensional, but
(2+1)-dimensional.  One has for the integration measure, whenever
the integrand is proportional to the gap
 \bea
 \int\frac{d^4\ell}{(2\pi)^4}&=&\int\frac{d\vec\ell_\perp}{(2\pi)^3}
 \int_{-\delta}^{+\delta}d\ell_\parallel\int_{-\infty}^{+\infty}\frac{d\ell_0}{2
\pi}=\cr &=&\frac{4\pi\mu^2}{(2\pi)^3}\frac{k_R}{2}
\int_{-\delta}^{+\delta}d\ell_\parallel\int_{-\infty}^{+\infty}\frac{d\ell_0}{2\pi}
\ , \eea which implements the condition (\ref{tetaq}).

Before closing this Section we note some differences between the
present calculation and the one presented in \cite{LOFF5}. In
\cite{LOFF5} we did not work in the simplifying approximation
$\alpha_z=0$ (see eq. (\ref{alfaz}), which implies that the
effective lagrangian and the fermion propagator in \cite{LOFF5}
contain both $2q$ and $\delta\mu$ and are therefore more involved
(there is also a numerical mistake in \cite{LOFF5} that renders
hermitean  the gap matrix $\Delta_{AB}$, which is not true. The
correct expression is eq. (\ref{eq:38})).
\section{Gap equation\label{2.7}}
To derive the gap equation in the CFL case we write a truncated
Schwinger-Dyson equation, similarly to the approach followed by
\cite{LOFF2001}. We assume
 a fictitious gluon propagator given by\be
i\,D^{\mu\nu}_{ab}=\,i\,\frac{g^{\mu\nu}\delta^{ab}}{\Lambda^2} \
,\ee which corresponds to a local four-fermion coupling. Using the
Feynman rules that can be derived by the effective theory one
obtains, in the limit of eq.(\ref{tetaq}) two equations: \bea
\Delta_0&=&
\,i\,\frac{\mu^2}{24\Lambda^2\pi^3}k_R\sum_{A=0}^3\,|\mu_A|\,I(\mu_A)
\cr&&\cr
 \Delta_1&=& - \,i\,\frac{\mu^2}{24\Lambda^2\pi^3}k_R\left(\mu_0I(\mu_0)+
 \mu_1 I(\mu_1)\right)\ ,\label{2.186}\eea
 where
\be I(\Delta)\,=\,\int \frac{d^2\ell}{V\cdot\ell\,\tilde
V\cdot\ell-\Delta^2+i\epsilon}= \,-\,i\,2\pi\,{\rm
arcsinh}\left(\frac\delta {|\Delta|}\right)\ . \ee

We fix $\Lambda$ by using the same equation and the same
fictitious gluon propagator at $\mu=0$, where the effect of the
order parameter, in this case the chiral condensate, is to produce
an effective mass $M$ for the light quark. This assumption gives
the equation \be
1=\frac{4}{3\Lambda^2\pi^2}\int_0^{K}dp\,\frac{p^2}{\sqrt{p^2+M^2}}\
. \ee For $M=400 $ MeV and the cutoff $K=800$ MeV (700 MeV) we get
$\Lambda= 181$ MeV ($154 $ MeV).

On the other hand we can consider the analogous equation for the
2SC model which is given by
 \be
1=\frac{i\,\mu^2}{3\Lambda^2\pi^3}\,I(\Delta)=\frac{2\,\mu^2}{3\Lambda^2\pi^2}\,
{\rm
arcsinh}\left( \frac{\delta}{|\Delta|}\right) \label{2sc}\ee that
can be solved explicitly: \be |\Delta|=\frac{\delta}{\dd
\sinh\left(\frac{3\Lambda^2\pi^2}{2\mu^2}\right)}\ .
\label{2scbis}\ee With the same values of $\Lambda$ ($\Lambda=
181$ MeV  and $154 $ MeV) we find the values for the gap parameter
of the 2SC model reported in the following table, where we have
taken $\delta$
of the order of $\mu$.\\
\begin{center}
\begin{tabular}{|c|c|}
\hline $\delta=K-\mu$ & $\Delta$  \\
\hline
  400\,MeV & 39 \,MeV \\
  300\,MeV & 68 \,MeV\\ \hline
\end{tabular}
\vskip0.5cm Table 1. {\it Value of the gap parameter $\Delta$ in
the 2SC model for two values of the cutoff $K$. The chemical
potential has the value $\mu=400$ MeV}.
\end{center}

Similar results have been obtained by \cite{alford} with a
procedure which differs from the present one for two reasons: They
include the ${\cal O} (1/\mu)$ corrections to the gap equation and
use a smooth cutoff instead of the sharp cutoff $K$ used here.

The coupled equations for the LOFF case (\ref{2.186}) have a non
trivial solution: \be \Delta_0=\frac{\delta}{\sinh
\left(\displaystyle{\frac{3\Lambda^2\pi^2}{\mu^2k_R}}\right)}\
,~~~~ \Delta_1=0\ .\label{68}\ee

This is the result for the gap in the LOFF phase in the
approximation $\mu\to\infty,\,q\to\infty\,$. It is known that the
crystalline phase holds in a well defined window of values
$|\delta\mu|$: $|\delta\mu|\in(\delta\mu_1,\,\delta\mu_2)$. The
determination of $\delta\mu_1$ is obtained by comparing the free
energies of the normal and 2SC states and is completely analogous
to the one in \cite{LOFF}, based on eq. (4.3) of this paper; one
gets, in the weak coupling limit $\delta\mu_1\approx
0.71\Delta_{2SC}$. On the other hand $\delta\mu_2$ is the limiting
value of $|\delta\mu|$ between the LOFF and the normal phase; at
$|\delta\mu|\approx\delta\mu_2$ one expects a second order phase
transition and  $\Delta_0\to 0$. In our approximation the phase
transition occurs because, as it is evident from (\ref{28}), for
sufficiently large $\delta\mu$ and fixed $q$ (in the analysis of
\cite{LOFF}, in the LOFF window $q$ is constant $\approx
0.90\Delta_{2SC}\approx 35.3$ MeV) $\cos\theta_q$ is not inside
the integration region in $\cos\theta$ and the only solution of
the gap equation is $\Delta_0=0$. Therefore, within this
approximation, $\delta\mu_2\simeq q\simeq 0.90\Delta_{2SC}$, which
compares favorably with the result $\approx 0.754\Delta_{2SC}$ of
\cite{LOFF}.

 Going beyond this approximation implies the use of the eq.
 (\ref{4bis}) alone, instead of  (\ref{4bis}) and (\ref{tetaq}); the  result
  (\ref{4bis}) is indeed more robust, based as it is
   only on the limit $\mu\to\infty$ and not also on
$q\to\infty$. Now the gap equation would assume the form
($\delta=\mu$):\bea
1&=&\frac{\mu^2\pi}{3\Lambda^2\pi^2R}\Big(\int_{-1}^{0}dz\,
\delta_R(h_-(z))\ {\rm arcsinh}\frac{\mu}{\Big|\frac \pi
R\delta_R(h_-(cos\theta)\,\Delta\Big|} \cr
&+&\int_{0}^{+1}dz\,\delta_R(h_+(z)){\rm arcsinh}\frac{\mu}{\Big |
\frac \pi R\delta_R(h_+(cos\theta)\,\Delta\Big |} \Big)\
.\label{beta}\eea Also for this approximation  we find  non
vanishing gaps; for example at $R\simeq 1$ and for
$\delta\mu=\delta\mu_1$, we get $\Delta_0\approx 5.5$ MeV; in this
case $\delta\mu_2\approx 1.25\Delta_{2SC}$; for $R\simeq 1.5$ we
find smaller values for the gap: $\Delta_0\approx 0.25$ MeV and
$\delta\mu_2\approx 0.83\Delta_{2SC}$. We observe explicitly that
in the limit of large $R$ one gets, as a solution of (\ref{beta}),
the  eq. (\ref{68}) with $\delta=\mu$.

 We finally notice that
we find $\Delta_1=0$, whereas in the more precise calculation of
ref. \cite{LOFF} it was found a value $\Delta_1\not=0\,;$ but very
suppressed with respect to $\Delta_0$. Presumably the difference
comes from our expansion at the leading order in $\mu$, and this
would explain the suppression of the vector condensate.

 Notice that, even though our
analysis leads to $\Delta_1=0$, we will consider in the sequel for
completeness the general case where both condensates are present.
\section{Phonon-quark interaction\label{sec5}}

Let us consider again the breaking terms in (\ref{ldeltaeff}),
i.e. ${\cal L}_\Delta^{(s)}$ and ${\cal L}_\Delta^{(v)}$.
 This
lagrangian explicitly breaks rotations and translations as it
induces a lattice structure given by parallel planes perpendicular
to $\vec n$:\be \vec n \cdot\vec x\,=\,\frac{\pi k}{q}\hskip 1cm
(k\,=0,\, \pm 1,\,\pm 2,...) \ .\label{planes}\ee We can give the
following physical picture of the lattice structure of the LOFF
phase: Due to the interaction with the medium,  the Majorana
masses of the red and green quarks oscillate in the direction
$\vec n$, reaching on subsequent planes maxima and minima. The
lattice planes  can fluctuate as follows:\be \vec n \cdot\vec
x\,\to\,\vec n \cdot\vec x\,+\,\frac{\phi}{2qf}\ ,\ee which
defines the phonon field, i.e. the Nambu-Goldstone boson that is
associated to the breaking of the space symmetries and has zero
vacuum expectation value: \be \langle
\phi\rangle_0=0\label{r22}\,.\ee Let us observe that, because of
(\ref{planes}), we have a fluctuating field $\phi_k$ for any nodal
plane, where \be
\phi_k\equiv\phi(t,x,y,z_k=k{\pi}/{q})\label{83}\ee Also the
vector £$\vec n$ can fluctuate: \be \vec n\,\to\,\vec R\
,\label{43} \ee where the
field vector $\vec R$ satisfies\bea |\vec R|&=&1\ ,\label{r1}\\
\langle\vec R\rangle_0&=&\vec n\label{r2}\,.\eea Also the vector
fields $\vec R$ represent, similarly to (\ref{83}), a collection
of fields: \be \vec R \equiv \vec R_k\ . \ee However we are
interested to give an an effective description of the fields
$\phi_k$ and $\vec R_k$ in the low energy limit, i.e. for
wavelengths much longer than the lattice spacing $\sim 1/q$; in
this limit the fields $\phi_k$ and $\vec R_k$ vary almost
continuously and can be imagined as continuous functions of three
space variables $x$, $y$ and $z$. Therefore we shall use in the
sequel the continuous notation $\phi$ and $\vec R$ and postpone a
discussion on this aspect to the end of the present Section and to
the subsequent Section.

The vector field $\vec R$ behaves as a vector under rotations
while being invariant under translations. If we take the $z$-axis
pointing along the direction $\vec n$, we can write \be (\vec
R)_i=R_{i}(\xi_1,\xi_2)=(e^{i(\xi_1 L_1+\xi_2
L_2)/f_R})_{i3}\label{campi_xi}\ee with $\vec L$ the generators of
the rotation group in the spin 1 representation, i.e. \be
(L_i)_{jk}=-i\epsilon_{ijk}\ee and $\xi_1$, $\xi_2$ are fields. At
the lowest order in these fields we find \be \vec R\approx \vec
n+\delta\vec n\label{48}\ee with \be\delta\vec n=
\left(-\frac{\xi_2}{f_R},\,+\frac{\xi_1}{f_R},\, 0\right)\
.\label{49}\ee Since the fields $R_{i}$ transforms as a vector
under rotations, it follows that the rotational symmetry is
restored by the substitution (\ref{43}).

If the direction $\vec q$ appearing in the exponent in
(\ref{ldeltaeff}) were different from the direction $\vec n$ of
the spin 1 condensate in the same equation, the physical situation
would not differ from magnetic materials; we would need a NGB to
take into account the fluctuations of  the scalar condensate
(\ref{scalar}), i.e. \be\Delta\,e^{2i\vec q\cdot\vec
x}\label{vec}\ee and other independent   fields $\delta\vec n$,
behaving as spin waves, i.e. making a precession motion around the
direction $\vec n$. Here, however, the situation is different,
because the two directions coincide. Therefore  the value of $\vec
R$ is strictly related to the symmetry breaking induced by the
scalar condensate.

We have shown in \cite{LOFF5} that in order to describe the
spontaneous breaking of space time symmetries induced by the
condensate (\ref{vec}) one NGB is sufficient. As we stressed in
\cite{LOFF5}, rotations and translations at least locally cannot
be seen as transformations breaking the symmetries of the theory
in an independent way, because the result of a translation plus a
rotation, at least locally,  can be made equivalent to a pure
translation \footnote{Recently the problem of the Goldstone
theorem for Lorentz invariant theories with spontaneous breaking
of  space-time symmetries has been considered in \cite{Low}.}.

 As a matter of fact at the lowest order we write
 \be
 \frac{\phi}{2qf}\,=\,(\vec n+\delta\vec n)\cdot(\vec x+
 \delta\vec x)-\vec n\cdot\vec x\approx\,\vec R\cdot\vec x+\frac T{2q}-\vec
n\cdot\vec x\ ,\label{35}\ee
 where we have introduced, as in \cite{LOFF5},  the auxiliary function $T$,
  given, in the present approximation, by
 \be T\,=\,2\,q\,\vec n\cdot\delta\vec x\ .\ee
Now the lattice fluctuations $\dd\frac\phi{2qf}$ must be small, as
the phonon is a longwavelength small amplitude fluctuation of the
order parameter. From (\ref{35}) it
  follows that $T$ must depend functionally on $\vec R$, i.e. $T=F[\vec R]$,
 which, using again (\ref{35}), means
  that\be \frac \Phi{2q}\equiv\vec n\cdot\vec x+\frac{\phi}{2qf}
 =\vec R\cdot\vec x+\frac{F[\vec R]}{2q}\equiv G[\vec R]\ .\ee
 The solution of this functional relation has the form
 \be \vec R=\vec
 h[\Phi]\ee where $\vec h$ is a vector built out of the scalar function
 $\Phi$. By this function one can only \footnote{In principle
 $\vec R$ could depend
linearly on a second vector, $\vec x$, but this possibility is
excluded for the assumed transformation properties under
translations.} form the
 vector $\vec\nabla \Phi$; therefore  we
 get
 \be
\vec R =\frac{\vec \nabla\Phi}{|\vec \nabla\Phi|}\, ,\label{r3}\ee
which satisfies (\ref{r1}) and, owing to (\ref{r22}), also
(\ref{r2}). In terms of the phonon field $\phi$ the vector field
$\vec R$ is given (up to the second order terms in $\phi$) by the
expression \be \vec R=\vec
n+\frac{1}{2fq}\left[\vec\nabla\phi-\vec n(\vec n\cdot\vec
\nabla\phi)\right]+\frac{\vec n}{8f^2 q^2}\left[3(\vec
n\cdot\vec\nabla\phi)^2-|\vec\nabla\phi|^2\right]-\frac{\vec\nabla\phi}{4f^2q^2}
(\vec n\cdot\vec\nabla\phi)\,.\label{63}\ee Note that at the
lowest order the fields $\xi_1$, $\xi_2$ introduced in (\ref{49})
are given by \bea
-\frac{\xi_2}{f_R}&=&\frac{1}{2fq}\frac{\partial\phi}{\partial x}
\, ,\cr
+\frac{\xi_1}{f_R}&=&\frac{1}{2fq}\frac{\partial\phi}{\partial y}
\ .\label{derivata}\eea  We observe that these relations do not
involve $\de\phi/\de z$. In fact we are defining the phonons as
fluctuations of the lattice planes $\dd z=k\frac{\pi}{q}$. In the
case of a single plane it is clear that the phonon propagates only
along the plane (say, $x$ and $y$ directions), and the only motion
along $z$ would be a translation corresponding to a zero mode.
However in our problem we have infinite Goldstone bosons
propagating on the planes $z=z_k=k{\pi}/{q}$. That is we have a
collection of fields as in eq. (\ref{83}). The presence of many
planes gives rise to a finite energy mode corresponding to the
variation of the relative distance among the planes. Phrased in a
different way, different phonon fields interact each other since
the fermions propagate in the whole three-dimensional space.
Translating the fields in momentum space we have to take a Fourier
series with respect to the discrete coordinate $z_k$. This
introduces in the theory a quasi-momentum $p_z$ through the
combination $\exp(ip_z z_k)$. Therefore $p_z $ and $p_z+2q$ define
the same physical momentum and we can restrict $p_z$ to the first
Brillouin zone $-q\le p_z\le q$. As already stressed. this point
will not affect our final discussion since we will be interested
to energies and momenta much smaller than the gap.

The interaction term with the NGB field   is contained in \be
{\cal L}_{int}=-\, e^{i\phi/f} \sum_{\vec
v}\left[\Delta^{(s)}\epsilon_{ij}+\Delta^{(v)}(\vec v\cdot\vec R
)\sigma^1_{ij} \right]\epsilon^{\alpha\beta 3}\psi_{i,\alpha,\vec
v}\,C\,\psi_{j,\beta,-\vec v}\,-\,(L\to R)\ +\ h.c.
\label{external}\ee Notice that we have neglected the breaking of
the color symmetry, which has been considered elsewhere
\cite{sannino}. At the first order in the fields one gets the
following   three-linear coupling:
 \bea
 {\cal L}_{\phi\psi\psi}&=&-\frac{i\phi}{f}
 \sum_{\vec
v}\left[\Delta^{(s)}\epsilon_{ij}+ \vec v\cdot\vec n
\Delta^{(v)}\sigma^1_{ij}\right]\epsilon^{\alpha\beta
3}\psi_{i,\alpha,\vec v}\,C\,\psi_{j,\beta,-\vec
v}\cr&&\cr&&-\frac{1}{2fq} \sum_{\vec v} \vec v\cdot \left[ \vec
\nabla\phi-\vec n(\vec n\cdot\vec\nabla\phi)\right]
\Delta^{(v)}\sigma^1_{ij}\epsilon^{\alpha\beta
3}\psi_{i,\alpha,\vec v}\,C\,\psi_{j,\beta,-\vec
v}\cr&&\cr&&-(L\to R)\ +\ h.c.
 \label{Trilineare}\eea
We  also write down the quadrilinear coupling:
 \bea
 {\cal L}_{\phi\phi\psi\psi}&=&\frac{\phi^2}{2f^2}\,
\sum_{\vec v}\left[\Delta^{(s)}\epsilon_{ij}+ \vec v\cdot\vec n
\Delta^{(v)}\sigma^1_{ij}\right]\epsilon^{\alpha\beta
3}\psi_{i,\alpha,\vec v}\,C\,\psi_{j,\beta,-\vec
v}\cr&&\cr&&-\frac{i\phi}{f}\, \sum_{\vec v} \vec v\cdot \left[
\vec \nabla\phi-\vec n(\vec n\cdot\vec\nabla\phi)\right]
\Delta^{(v)}\sigma^1_{ij}\epsilon^{\alpha\beta
3}\psi_{i,\alpha,\vec v}\,C\,\psi_{j,\beta,-\vec v}\cr&&\cr&&
-\frac{ 1}{4f^2q^2}\,\sum_{\vec v} \left[\frac {\vec v\cdot \vec
n} 2 \left(3(\vec
n\cdot\vec\nabla\phi)^2-|\vec\nabla\phi|^2\right)- (\vec
v\cdot\vec \nabla\phi)(\vec n\cdot\vec\nabla\phi)\right]\times
\cr&&\cr&&\times \Delta^{(v)}\sigma^1_{ij}\epsilon^{\alpha\beta
3}\psi_{i,\alpha,\vec v}\,C\,\psi_{j,\beta,-\vec v}   -(L\to R)\
+\ h.c.
 \label{quadrilineare}\eea
We notice that in (\ref{external}), (\ref{Trilineare})and
(\ref{quadrilineare}) the average over the velocities is defined
according to (\ref{sum3}).

It is now straightforward to re-write these couplings in the basis
of the $\chi$ fields. One gets
 \be {\mathcal L}_3\,+\,{\mathcal L}_4=
 \sum_{\vec v}\sum_{A=0}^3
\tilde\chi^{A\,\dag}\,   \left(
\begin{array}{cc}
 0 & g_3^\dag\,+\,g_4^\dag
\\
  g_3\,+\,g_4& 0\end{array}\right)
    \,\tilde\chi^B\ ,\label{vertex}\ee
Here\bea g_3&=&\,\left[\frac{i\phi\Delta_{AB}}{f}+\sigma_{AB}\hat
O[\phi] \right] \ ,\cr
g_4&=&\,\left[-\frac{\phi^2\Delta_{AB}}{2f^2}+\sigma_{AB}
 \left(\frac{i\phi}{f}\hat O[\phi]
 \,+\,\hat Q[\phi]\right)\right] \ ,\eea with \bea
 \hat O[\phi]&=&\frac{1}{2fq}\vec v\cdot \left[ \vec \nabla\phi-\vec n(\vec
n\cdot\vec\nabla\phi)\right] \Delta^{(v)}\ ,\cr
 \hat Q[\phi]&=&
\frac{\Delta^{(v)}}{4f^2q^2}\left[\frac {\vec v\cdot\vec n}
2\left(3 (\vec n\cdot\vec\nabla\phi)^2-|\vec\nabla\phi|^2\right)-
(\vec v\cdot\vec \nabla\phi)(\vec n\cdot\vec\nabla\phi)\right]\ ,
 \eea
\bea \sigma_{AB}&=& \left(\begin{array}{cccc}
  0 & 0 & 0 & -1 \\
  0 & 0 & -i & 0 \\
  0 & +i & 0& 0 \\
 +1 & 0 & 0 & 0
\end{array}\right)\ .
\eea For the following it is important to observe that the term in
$g_3$ and $g_4$ proportional to $\Delta_{AB}$ arise from the
expansion of $\exp{i\phi/f}$ alone, whereas the terms proportional
to $\sigma_{AB}$ get also contribution from the expansion of $\vec
R$ in the vector condensate.

 Before concluding this section, let us come back to the
lattice structure given by the parallel planes of Eq.
(\ref{planes}). The effective action for the field $\phi$, $S[
\phi ]$, is obtained by the lagrangian  as follows \be S=\int
dt\,dx\,dy\ \frac{\pi}{q}\sum_{k=-\infty}^{+\infty}{\cal
L}(\phi(t,x,y,k\pi/q)\ ,\ee In the action bilinear terms of the
type $\phi_k\phi_{k^\prime}$ with $k\neq k^\prime$ may arise. In
the continuum limit this terms would correspond to derivatives
with respect to the $z$ direction

We can use the same lagrangian ${\cal L}_3+{\cal L}_4$ given in
(\ref{vertex}) with the integration measure $\dd\int d^4 x$
provided we multiply the r.h.s. of (\ref{vertex}) by the factor
\be\tau(z)= \frac \pi q
\sum_{k=-\infty}^{+\infty}\delta\left(z-\frac{k\pi}{q}\right)\
.\label{tau}\ee As already  observed  the phonon momentum is a
quasi-momentum, i.e. $p_z$ and $p_z+2q$ correspond to the same
$z-$component of the momentum. Due to this remark one can make use
in the calculations of the
formula\be\sum_{k=-\infty}^{+\infty}e^{ik\pi
p/q}=\delta\left({\frac p{2q}}\right)\ .\ee

\section{Effective lagrangian for the phonon field\label{sec6}}
To introduce formally the NGB in the theory one can use the
gradient expansion (see e.g. \cite{eguchi}), i.e. a bosonization
procedure similar to the one employed in \cite{gatto} to describe
the goldstones associated to the breaking of $ SU(3)$ in the CFL
model.  First one introduces an external field with the same
quantum numbers of the NGB and then performs a derivative
expansion of the generating functional. This gives rise to the
effective action for the NGB. At the lowest order one has to
consider the diagrams in Fig. 1, i.e. the self-energy, Fig. 1a,
and the tadpole, Fig. 1b, whose result we name $\Pi(p)_{s.e.}$ and
$\Pi(p)_{tad}$ respectively.
\begin{figure}
\epsfxsize=8truecm \centerline{\epsffile{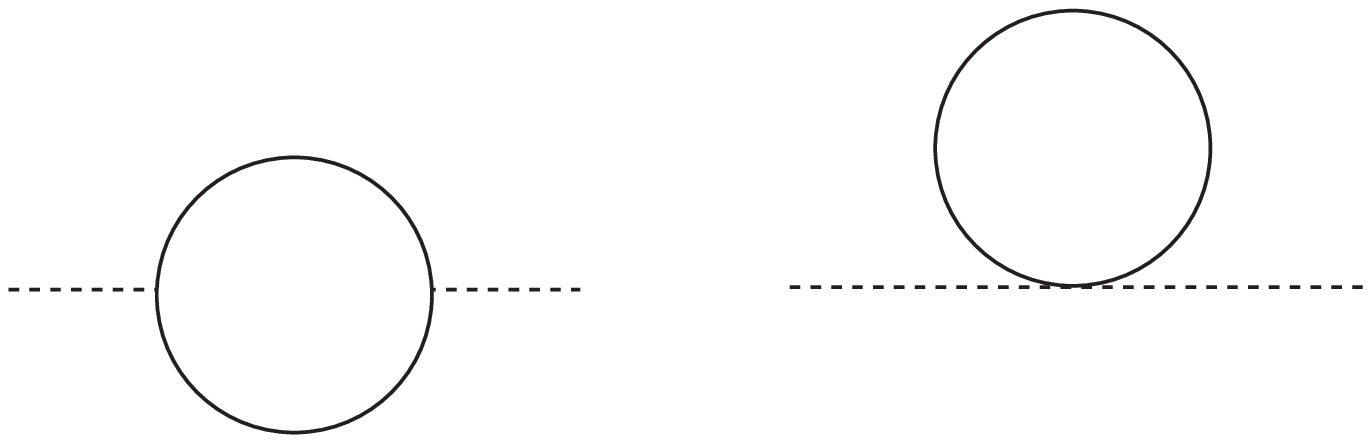}} \noindent
\hskip 4.5cm a) \hskip 4cm b) \par\noindent \vskip0.5cm Fig. 1.
Self-energy (a) and tadpole (b) diagrams.
\end{figure}

To  perform the calculation one employs the propagator given in
Eq. (\ref{propagatore}) and the interaction vertices in
(\ref{vertex}). The result of the calculation of the  two diagrams
at the second order in the momentum expansion is as follows: \bea
\Pi(p)_{s.e.}&=&
 \frac{i\,\mu^2}{16\pi^3f^2} \sum_{\vec
v}\sum_{C=0}^3\int  {d^2\ell} \Big[\frac{4\,\lambda_C^2}{
D^2_C(\ell)}\,-\,\frac{4\lambda_C\,V\cdot\ell\,\tilde V\cdot\ell
}{D_C^2(\ell)}
 \cr
 &&
 \cr \,
 &-&\,
 4\lambda^2_C\frac{V\cdot p\,\tilde
V\cdot p}{D_C^3(\ell)}\, -\,\left(\frac{\Delta^{(\vec v
)}}{q}\right)^2\omega^2(\vec p)\left(
\frac{2\lambda_C}{D^2_C(\ell)}\,+\,\frac{1}{D_C(\ell)}
\right)\Big]\ ,
 \cr
 &&
 \cr
 \Pi(p)_{tad}
 &=&
 \frac{i\,\mu^2}{16\pi^3f^2}
  \sum_{\vec v}\sum_{C=0}^3
  \int  \frac
{d^2\ell}{D_C(\ell)}\,\Big[\,4\,\lambda_C \,-\,\frac{\Delta^{(\vec
v )}}{q^2}\,\mu_C\,\times \cr&\times&
\left(-p_x^2-p_y^2\,+2p_z^2-\,2\vec p\cdot\vec v\,p_z \right)\Big]
 \eea
  where
  \be
D_C(\ell)=\ell_0^2-\ell_\parallel^2-\lambda_C+i\epsilon\hskip0.3cm
, \ee $\mu_c$ defined in (\ref{muc}) and \be
 \omega(\vec
p)=\vec p\cdot\vec v-(\vec p\cdot\vec n)(\vec v\cdot\vec n)
 \ .
\ee
 One can easily control that the Goldstone theorem is
satisfied and the phonon is massless because one has
\be\Pi(0)=\Pi(0)_{s.e.}\,+\,\Pi(0)_{tad}\, =\,0\ .\ee In
performing the integration we take into account the circumstance
that $\delta\gg |\mu_C|$.  At the second order in the momentum
expansion one gets
 \be\Pi(p)\,=\,-\,\frac{\mu^2}{2\pi^2 f^2}\sum_{\vec v}
 \Big[
 V_\mu\tilde V_\nu p^\mu p^\nu+\Omega^{(v)}(\vec p)\Big]\ . \ee
 Here
 \bea\Omega^{(v)}(\vec p)&=&
 -\,\left(\frac{\Delta^{(\vec v
)}}{q}\right)^2\omega^2(\vec p)\left(2-\frac 1 2 \sum_{C=0}^3{\rm
arcsinh}\frac\delta{|\mu_c|}\right)\,+\cr&&\cr&+&
\frac{\Delta^{(v)}}{2q^2} \Phi(\vec p) \sum_{C=0}^3\mu_C\times{\rm
arcsinh}\frac\delta{|\mu_c|}
 \,\approx\cr&&\cr&\approx&
-2\,\left(\frac{\Delta^{(\vec v
)}}{q}\right)^2\left(1-\log\frac{2\delta}{\Delta_0}\right)\left(\omega^2(\vec
p)+\vec v\cdot\vec n\Phi(\vec p)\right) \label{final2}\eea where
we have taken $\Delta^{(v)}\ll\Delta_0$ and \be \Phi(\vec
p)=\left(3p_z^2-{\vec p}^{\,2}\right)\vec v\cdot\vec n \,-\,2\vec
p\cdot\vec v\,p_z \ee  After averaging over the fermi velocities
we obtain \be \Pi(p)\,=\,-\,\frac{\mu^2k_R}{4\pi^2 f^2}\Big[
 p_0^2-v_\perp^2(p_x^2+p^2_y)-v_\parallel^2 p^2_z
 \Big]\, .\label{final2bis}\ee
 One obtains canonical normalization for the kinetic term
provided \be
 f^2= \frac {\mu^2k_R}{2\pi^2}
  \ .
\ee  On the other hand
 \bea v_\perp^2&=&\,\left[\frac 1
 2\sin^2\theta_q+\left(1-3\cos^2\theta_q\right)\left(1-\log\frac{2\delta}{\Delta
_0}\right)
 \left(\frac{\Delta^{(v)}}q\right)^2\right]
 \\ &&\cr
 v_\parallel^2&=&\,\cos^2\theta_q
  \ . \eea
  We note that $f^2$ depends on $k_R$. We
expect that $k_R$ is in the range (1,\,2), where  1 corresponds to
a smearing distance equal to the lattice spacing $|\Delta
z|=\pi/q$, and 2 corresponds to  realistic values for the gap. It
is remarkable however that the phonon dispersion relation is not
affected by this uncertainty. We also notice that the only source
of anisotropy is in the asymmetric average over the Fermi
velocity, that is the integration over a circle instead of the
integration over a sphere. In fact, if we were going to average
the previous result over $\cos\theta_q$ we would get the isotropic
result $v_\perp^2=v_\parallel^2=1/3$. In particular the
contribution from ${\Delta^{(v)\,2}}$ averages to zero. In fact,
this contribution is just the one arising from the terms in $g_3$
and $g_4$ proportional to $\sigma_{AB}$. For the sake of the
argument let us forget for a moment the relation between $\Phi$
and $\vec R$. In this case the previous contribution is nothing
but the "zero-momentum" contribution to the two point function for
the fields $\xi_i$ (or $\vec\delta n$) of eq. (\ref{campi_xi}). In
the actual calculation this is a second order term in the momenta
simply because the fields $\xi_i$ are derivatives of $\phi$. It is
a simple consequence of the symmetries to see that, after
averaging the velocity over all the Fermi sphere, this term is
necessarily proportional to $\xi_1^2+\xi_2^2\approx
(\vec\delta\vec n)^2$. In fact, the only vectors involved are
$\delta\vec n$, $\vec n$ and $\vec v$. Since $\delta\vec
n\cdot\vec n=0$, after averaging  the result is necessarily
proportional to $(\delta\vec n )^2$. However, the Goldstone
theorem tells us that Goldstone fields should have no mass. This
is true also for $\vec\delta n$, since being a gradient of $\phi$
it still satisfies the same wave equation of $\phi$. This explains
the particular expression for this term, which has to average to
zero integrating the velocity over all the Fermi sphere.

In any event the term proportional to the vector condensate
$\Delta^{(v)}$ is very small and the result for $v_\perp^2$ can be
approximated as follows
 \be
  v_\perp^2\simeq\frac{1}2\,\sin^2\theta_q
  \ . \ee

In conclusion, the dispersion law for the phonon is \be E(\vec
p)=\sqrt{v_\perp^2(p_x^2+p^2_y)+v_\parallel^2 p^2_z}\ \ee which is
anisotropic. Besides the anisotropy related to
 $v_\perp\neq v_\parallel$, there is another source of
 anisotropy,  due to the fact that
 $p_z$, the component
 of the momentum perpendicular to the planes (\ref{planes}),
 differently from $p_x$ and $p_y$
 is a quasi momentum and not a real momentum. The difference can
 be better  appreciated in coordinate space, where the effective
 lagrangian reads
  \be {\cal L}=\frac 1 2\left[(\dot\phi_k)^2-v_\perp^2
  (\de_x\phi_k)^2-v_\perp^2 (\de_y\phi_k)^2-v_\parallel^2\left(\frac
q\pi\right)^2(\phi_k-\phi_{k-1})^2\right]\ .\ee However, in the
long distance limit $\ell>>\pi/q$, the set of fields $\phi_k(x,y)$
becomes a  function $\phi(x,y,z)$ and the last term can be
approximated by $v_\parallel^2(\de_z\phi)^2$.

\section{Conclusions\label{secconc}}

The formalism we have used employs velocity dependent fields and
it has been adapted to describe anisotropic phases, such as the
crystalline phase of interest here. We have considered QCD with
two massless flavors. For the two-quark condensates ( scalar and
vector) we have made the currently used plane-wave ansatz. For
different values of the chemical potentials of up- and down-quark
the two fermion momenta require independent decompositions. The
quark propagators can be derived from the effective action for the
fermi fields, as provided by the velocity dependent formalism,
neglecting antiparticles (with respect to the Fermi sphere). A
careful discussion is required for the integration over
velocities, due to the anisotropy of the LOFF phase, implying that
the effective theory is (2+1)-dimensional, differently from the
situation that applied to CFL or to 2SC where it was
(1+1)-dimensional.

We have derived the gap equation for the normal BCS state and the
LOFF phase within the same formalism. Our approach is based on a
few approximations, most notably the neglecting of the negative
energy states and subleading terms in the $\mu\to\infty$ limit.
These approximations considerably simplify the formalism and
render it suitable for more complicated computations such as those
arising from crystalline patterns more involved than the simple
plane wave ansatz assumed in this paper. In spite of this
approximations we are able to obtain  results that qualitatively
agree with more complete calculations existing in the literature.

The breaking of rotations and translations induces within the
plane-wave ansatz a lattice structure of parallel planes
corresponding to maxima and minima of the fermion Majorana masses.
Fluctuations of the lattice planes define the phonon field. The
phonon momentum has to be regarded as a quasi-momentum, but one
can restrict to the first Brillouin zone. The formal introduction
of the Nambu-Goldstone boson has been performed through
bosonization, leading to the effective boson action.

The main result of the present paper is the application of the
formalism to derive the parameters of the phonon field, i.e. the
goldstone field associated to the breaking of the rotational and
translational invariance in the crystalline color superconductive
phase of QCD. Lowest order calculation requires self-energy and
tadpole diagrams. They have to be calculated in terms of the
fermion propagators and the three-linear and quadrilinear
phonon-quark interactions. Besides explicit verification of the
zero value of the phonon mass, we have derived the values of the
phonon constant $f$ and the expressions for the transverse and
parallel velocities. The resulting phonon dispersion relation is
anisotropic, both because of the difference in the velocity
components, and of the quasi-momentum aspect of one momentum
component. In coordinate space, the effective lagrangian exhibits
explicitly such anisotropy in a most apparent form.


\begin{thebibliography}{1-99}
%\bibitem{alford}M. Alford, K. Rajagopal and F. Wilczek, Phys.
%Lett.B {\bf 422} (1998) 247.
\bibitem{review} For a review see K.Rajagopal and F.Wilczek,
in Handbook of QCD, M.Shifman ed. (World Scientific 2001), hep-ph/
0011333. For earliest papers on color superconductivity see B.
Barrois, Nucl. Phys. B {\bf 129}, 390 (1977); S. Frautschi,
Proceedings of workshop on hadronic matter at extreme density,
Erice 1978. See also: D. Bailin and A. Love, Phys. Rept. {\bf
107}, 325 (1984) and references therein.
\bibitem{LOFF}M. Alford, J. A. Bowers and K. Rajagopal,
Phys.Rev. D {\bf 63}, 074016 (2001)  hep-ph/0008208.
%
\bibitem{LOFF2001}J. A. Bowers, J. Kundu, K. Rajagopal and E.
Shuster, Phys. Rev. D {\bf 64},  014024 (2001) hep-ph/0101067.
%
\bibitem{LOFF5}
R.Casalbuoni, R. Gatto, M. Mannarelli and G. Nardulli, Phys. Lett.
B {\bf 511}, 218 (2001) hep-ph/0101326.
%
\bibitem{LOFFbis}A. K. Leibovich, K. Rajagopal and E. Shuster,
 Phys. Rev. D {\bf 64}, 094005 (2001) hep-ph/0104073.
%
\bibitem{LOFF7}
D. K. Hong, Y. J. Sohn, hep-ph/0107003.

\bibitem{LOFF8}
K. Rajagopal,  in the proceedings of QCD@Work: International
Workshop on  QCD: Theory and Experiment, Martina Franca, Bari,
Italy,  2001, editors P. Colangelo and G. Nardulli, AIP Conference
Proceedings, V. 602, 2001, hep-ph/0109135.

\bibitem{LOFF_rob}
 R Casalbuoni, in the proceedings of QCD@Work: International
Workshop on  QCD: Theory and Experiment, Martina Franca, Bari,
Italy,  2001, editors P. Colangelo and G. Nardulli, AIP Conference
Proceedings, V. 602, 2001, hep-th/0108195.

\bibitem{LOFF9}
G. Nardulli, Invited talk at Compact Stars in the QCD Phase
Diagram, Copenhagen, Aug. 2001, hep-ph/0111178

\bibitem{LOFF10}
D. K. Hong, hep-ph/0112028.

\bibitem{massdiff}
J. Kundu and K. Rajagopal, hep-ph/0112206.
%
\bibitem{LOFF3} K. Rajagopal, Acta Phys. Polon. B {\bf 31}, 3021 (2000)
hep-ph/0009058;
 M. Alford, J. A. Bowers and K. Rajagopal, J. Phys.
G {\bf 27}, 541 (2001) hep-ph/0009357; T. Sch\"afer and E.
Shuryak, nucl-th/0010049.
\bibitem{LOFF2}A. I. Larkin and Yu. N. Ovchinnikov, Zh. Eksp.
Teor. Fiz. {\bf 47}, 1964 (1136) (Sov. Phys. JETP {\bf 20}, 762
(1965)); P.Fulde and R. A. Ferrell, Phys. Rev. {\bf 135}, A550
(1964).
%
\bibitem{condmat}H. Shimahara, cond-mat/0203109.
%
\bibitem{nuclearmatter}A. Sedrakian, Phys. Rev. C {63}, 025801 (2001)
nucl-th/0008052.
%
\bibitem{coldtrapped}R. Combescot, cond-mat/0007191.
%
\bibitem{hong} D.~K. ~Hong, Phys. Lett. B {\bf
473}, 118 (2000) hep-ph/9812510; D.~K.~ Hong, Nucl. Phys. B {\bf
582}, 451 (2000) hep-ph/9905523.
%
\bibitem{beane} S. R. Beane, P. F. Bedaque and M.
J. Savage, Phys. Lett. B {\bf 483}, 131 (2000) hep-ph/0002209.
%
\bibitem{gatto}R. Casalbuoni, R. Gatto and G. Nardulli,  Phys. Lett.
B {\bf 498}, 179 (2001) hep-ph/0010321.
%
\bibitem{mannarelli}R. Casalbuoni, R. Gatto, M. Mannarelli
and G. Nardulli, Phys.Lett.B {\bf524}, 144 (2002).
 hep-ph/0107024.
%
\bibitem{alford}M. Alford, K. Rajagopal and F. Wilczek, Phys.
Lett. B {\bf 422}, 247 (1998).
%
\bibitem{Low}
I. Low and A.V. Manohar, hep-th/0110285.
%
\bibitem{sannino}R. Casalbuoni, Z. Duan and F. Sannino,
Phys. Rev. D {\bf 62},  094004 (2000) hep-ph/0004207.

\bibitem{eguchi}
T. Eguchi, Phys. Rev. D {\bf 14}, 2755 (1976).
\end{thebibliography}
\end{document}